\documentstyle[12pt,aasms4,epsf]{article}
\newcommand{\gapr}{\raisebox{-.6ex}{\mbox{
$\stackrel{>}{\mbox{\scriptsize$\sim$}}\:$}}}
\newcommand{\lapr}{\raisebox{-.6ex}{\mbox{
$\stackrel{<}{\mbox{\scriptsize$\sim$}}\:$}}}
\def\Tpc{T_{\rm pc}}
\def\Rpc{R_{\rm pc}}
\def\psr{PSR J0437--4715}
\def\fp{f_{\rm p}}
\def\R10{R_{\rm 10}}
\def\Ms{M_\odot}
\def\mr{M_*/\R10}
\begin{document}
\lefthead{Pavlov and Zavlin}
\righthead{Mass and Radius of \psr}
\title{Mass-to-Radius Ratio for the Millisecond Pulsar J0437--4715}

\author{G.~G. Pavlov}
\affil{The Pennsylvania State University, 525 Davey Lab,
University Park, PA 16802, USA; pavlov@astro.psu.edu}
\and
\author{V.~E. Zavlin}
\affil{Max-Planck-Institut f\"ur Extraterrestrische Physik, D-85740
Garching, Germany; zavlin@rosat.mpe-garching.mpg.de}
\begin{abstract}
Properties of X-ray radiation emitted from the polar caps 
of a radio pulsar depend not only on the cap
temperature, size, and position, but also on the surface chemical composition,
magnetic field, and neutron star's mass and radius. 
Fitting the spectra and the light curves 
with neutron star atmosphere models enables one to infer these parameters.
As an example, we present here results obtained from the analysis
of the pulsed X-ray radiation of a nearby millisecond 
pulsar J0437--4715. In particular, 
we show that stringent constraints on the mass-to-radius ratio
can be obtained if orientations of the magnetic and rotation axes
are known, e.g., from the radio polarization data.
\end{abstract}
\section{Introduction}
Virtually all of the different
models of radio pulsars (e.~g., Cheng \& Ruderman 1980;
Arons 1981; Michel 1991; Beskin, Gurevich \& Istomin 1993)
predict a common phenomenon: the presence of polar caps (PCs) 
around the neutron star (NS) magnetic poles heated 
up to X-ray temperatures by the backward accretion of relativistic particles 
and gamma-quanta from the pulsar magnetosphere.
A typical size of the PC is estimated to be close to the radius 
within which the open magnetic field lines originate from the
NS surface,
$\Rpc\sim (2\pi R^3/Pc)^{1/2}~(\sim 0.1 - 3$~km for the period
$P\sim 2~{\rm s}-2$~ms).  However, expected PC temperatures, 
$\Tpc\sim 5\times 10^5 - 5\times 10^6$~K,
and luminosities, $L_{\rm pc} \sim 10^{28}-10^{32}$~erg~s$^{-1}$,
are much less certain and strongly depend on the specific pulsar model.
Studying X-ray radiation from the PCs is particularly useful
to discriminate between different models. 

The best candidates for the investigation of the PC radiation 
are nearby, old pulsars of ages $\tau\gapr 10^6$~yr, 
including very old millisecond pulsars, for which 
the NS surface outside the PCs is expected
to be so cold, $T\lapr 10^5$~K, that its thermal radiation is
negligibly faint in the soft X-ray range.
Indeed, available observational data allow one to assume the PC 
origin of soft X-rays detected from, e.~g.,
PSR B1929+10 (Yancopoulos, Hamilton \& Helfand 1994; Wang \& Halpern 1997),
B0950+08 (Manning \& Willmore 1994; Wang \& Halpern 1997),
and J0437--4715 (Becker \& Tr\"umper 1993).
Moreover, there are some indications that hard tails of 
the X-ray spectra of younger pulsars, B0656+14 and B1055--52, may
contain a thermal PC component (Greiveldinger et al.~1996). 
On the other hand, nonthermal (e.~g., magnetospheric)
X-ray radiation may dominate even in very old pulsars
(cf. Becker \& Tr\"umper 1997). For example, $ASCA$ observations
of the millisecond pulsar B1821--24 (Saito et al.~1997),
whose luminosity in the 0.5--10~keV range
exceeds that predicted by PC models by a few orders of magnitude
(see discussion in Zavlin \& Pavlov 1997; hereafter ZP97),
proved its X-rays to be of a magnetospheric origin.
Thus, thorough investigations are needed in each specific case to
ascertain the nature of the observed radiation.

The closest known millisecond pulsar J0437--4715 
($P=5.75$~ms, $\tau = P/2\dot{P} = 5\times 10^9$~yr, 
$\dot{E}=4\times 10^{33}$~erg~s$^{-1}$, $B\sim 3\times 10^8$~G, $d=180$~pc) 
is of special interest.
Important data from this pulsar have been collected
with the $ROSAT$ (Becker \& Tr\"umper 1993;
Becker et al.~1997), $EUVE$ (Edelstein, Foster \& Bowyer 1995;
Halpern, Martin \& Marshall 1996) and $ASCA$ (Kawai, Tamura
\& Saito 1996) space observatories. 
Becker \& Tr\"umper (1993) and Halpern et al.~(1996)
showed that the spectral data can be fitted with power-law or
blackbody plus power-law models. The latter fit 
indicates that the X-ray emission may be, at least partly,
of a thermal (PC) origin.  Rajagopal \& Romani (1996)
applied more realistic models of radiation emitted by NS
atmospheres to fitting the thermal component and concluded
that the iron atmospheres do not fit the observed spectrum.

The observations made with the $ROSAT$ and $EUVE$ missions  
revealed smooth pulsations of soft X-rays
with the pulsed fraction $\fp\sim 25-50\%$ apparently
growing with photon energy in the 0.1--2.4~keV $ROSAT$ range. 
Since the pulsed fraction of the nonthermal
radiation is not expected to vary significantly in the 
narrow energy range, it is natural to attribute the observed radiation
to the pulsar PCs. 
The PC radiation should be inevitably pulsed unless the rotation axis
coincides with either the line of sight or magnetic axis.
If it were the blackbody (isotropic) radiation, 
the pulsed fraction would remain the same at all photon energies.  
The energy dependence of $\fp$ can be caused by 
anisotropy (limb-darkening) of thermal
radiation emitted from NS atmospheres (Pavlov et al.~1994; 
Zavlin, Pavlov \& Shibanov 1996).
Zavlin et al.~(1996) showed that even in the case of low magnetic fields 
characteristic to millisecond pulsars ($B\sim 10^8 - 10^9$~G) 
the anisotropy strongly 
depends on photon energy and chemical composition of NS surface layers.
To interpret the $ROSAT$ and $EUVE$ observations of \psr,
ZP97 applied the NS atmosphere models with
account for the energy-dependent limb-darkening
and the effects of gravitational redshift
and bending of photon trajectories (Zavlin, Shibanov \& Pavlov 1995). 
Assuming the viewing angle (between the rotation axis and line
of sight), $\zeta=40^\circ$, 
and the magnetic inclination (angle between the magnetic and
rotation axes), $\alpha=35^\circ$, inferred by Manchester \& Johnston (1995)
from the phase dependence of the position angle of the radio
polarization, ZP97 showed that 
both the spectra and the light curves (pulse profiles) of
the {\em entire} soft X-ray radiation detected by $ROSAT$ and $EUVE$ 
can be interpreted as thermal radiation
from two hydrogen-covered PCs, whereas 
neither the blackbody nor iron atmosphere models fit the observations.
The approach of ZP97 differs substantially from that of 
Rajagopal \& Romani (1996) who
did not take into account the energy-dependent anisotropy of
the emergent radiation and gravitational bending (hence, they
could not analyze the pulse profiles), made the unrealistic assumption
that pulsed and unpulsed flux components are of completely separate
origin, and consequently obtained quite different parameters
of the radiating region. 
The simplest, single-temperature PC model of ZP97
provides a satisfactory fit with typical PC radius $\Rpc\sim 1$~km
(reasonably close to the theoretical estimate of $1.9$~km) 
and temperature $\Tpc\sim 1\times 10^6$~K. 
The corresponding bolometric luminosity of the two PCs, 
$L_{\rm bol} = (1.0-1.6)\times 10^{30}$~erg~s$^{-1}$, comprises
$\sim (2-4)\times 10^{-4}$ of the pulsar total energy loss
$\dot{E}$.  This value of $L_{\rm bol}$ is in excellent agreement 
with the predictions of the slot-gap pulsar model by Arons (1981).
Even better fit to the observational data is provided by a model
with non-uniform temperature distribution along
the PC surface.  In addition, the inferred interstellar
hydrogen column density
towards the \psr, $n_H\sim (1-3)\times 10^{19}$~cm$^{-2}$
was demonstrated to be well consistent with the ISM properties obtained
from observations of other stars in the vicinity of the pulsar.
All these results allow one to 
conclude that the X-ray radiation observed from \psr~ is indeed of the
thermal (PC) origin. 

The results of ZP97 were obtained for fixed orientations of
the rotation and magnetic axes
(angles $\zeta$ and $\alpha$) and for standard NS mass
$M=1.4 \Ms$ and radius $R=10$~km. 
The inferred PC radius, temperature and luminosity are almost insensitive 
to these four parameters. Their main effect is on the shape and pulsed
fraction of the light curves. For instance, $\fp$ decreases
with increasing the mass-to-radius ratio, 
unless the observer can see the PC in the center of the back hemisphere
of the NS, which is possible at $\alpha \simeq \zeta$ and
$\mr > 1.93$, where
$M_*=M/\Ms$ and $\R10=R/(10$~km$)$ (e.~g., Zavlin et al.~1995). 
The angles $\alpha$ and $\zeta$ cannot be precisely evaluated
from the radio polarization measurements because of
the complicated variation of the polarization position angle
across the eight-component mean radio pulse
of \psr~(Manchester \& Johnston 1995),
and the true NS mass and radius may differ from the canonical values.
On the other hand, the fact that the set of angles 
and $M/R$ adopted by ZP97 fits the data does not mean that a better
fit cannot be obtained for another set, and it tells us nothing
about the allowed domain of these parameters. 
Hence, fitting the light curves for variable $M/R$, 
$\alpha$, and $\zeta$ enables one
to constrain the mass-to-radius ratio and/or the magnetic inclination
and viewing angle, so that the present paper is complementary to ZP97.
We describe our approach in \S 2 and present the results on \psr~in \S 3.
\section{Method}
Since the shape of the light curves depends
on energy, the light curve fitting is coupled to the spectral fitting.
We fit the count rate spectrum (total $\simeq 3200$ counts) collected by 
the $ROSAT$ Position Sensitive Proportional Counter (PSPC)
with the phase-integrated model spectrum emitted from two 
identical, uniformly heated PCs $180^\circ$ apart, 
assuming the hydrogen composition of the surface layers (ZP97).
The fitting is carried out on a grid of angles $\zeta$ and $\alpha$ 
between $0^\circ$ and $90^\circ$ at different mass-to-radius ratios
$\mr$ in a range allowed by equations of state of the superdense NS matter. 
As shown in ZP97, both the $ROSAT$ and $EUVE$ observations of \psr~
are consistent with  
the applied model at the interstellar hydrogen column density of 
$\sim 1\times 10^{19}$~cm$^{-2}$; therefore we freeze $n_H$ at this value
and obtain $\Tpc$ and $\Rpc$ from the spectral fits
for each set of $\zeta$, $\alpha$ and $\mr$.
With these $\Tpc$ and $\Rpc$, we calculate the model spectral fluxes
for various phases of the pulsar period and fold each of
the spectra with the PSPC response matrix;
this gives us the model light curve as a function
of phase $\phi$ for given $\zeta$, $\alpha$ and $\mr$.
This light curve is then compared with the observed PSPC
light curve. For this comparison, we used the PSPC light curve
for the total $ROSAT$ energy range, 0.1--2.4~keV; its pulsed
fraction (for 17 phase bins) was determined to be $\fp=30\pm 4\%$
(ZP97).  We bin the model light curve
to the same number of phase bins ($K=17$) and calculate the $\chi^2$ value,
\begin{equation}
\chi^2=\sum_{k=1}^{K} \frac{(N_{k, \rm o} - N_{k, \rm m})^2}
{N_{k,\rm o}}~,
\end{equation}
for each set of $\zeta$, $\alpha$ and $\mr$ ($N_{k,\rm o}$
and $N_{k,\rm m}$ are the observed and model numbers of counts
in the $k$-th bin). This allows us to find the best-fit parameters
(which correspond to the minimum of $\chi^2$) and confidence levels
in the parameter space. 
\section{Results}
Figure~1 shows the 68\%, 90\% and 99\% confidence regions 
for the above-described  light
curve models in the $\zeta$-$\alpha$ plane at several
values of the mass-to-radius ratio, 
$\mr=1.1, 1,2,\ldots 1.6$. The ratio determines the
parameter $g_r=\sqrt{1 - 0.295\mr}$
responsible for the effects of gravitational redshift and bending
of photon trajectories (e.~g., Zavlin et al.~1995).
For the $\mr$ values in Figure 1, the parameter $g_r$ varies
between 0.82 and 0.73, making visible from 74\% to 91\% of 
the whole NS surface, so that
a distant observer can detect the radiation from both PCs
simultaneously during almost the whole pulsar period.
The plots in Figure 1 are clearly  symmetrical with respect
to the transformation $\zeta\leftrightarrow \alpha$ because
the model light curves depend only on 
the angle $\theta$ between
the observer's direction and the magnetic axis:
$\cos\theta = \cos\zeta~\cos\alpha + \sin\zeta~\sin\alpha~\cos\phi$,
where $\phi$ is the rotational phase.
The minimum value of the reduced $\chi^2$ ($\chi^2_\nu=1.07$ for 
17 degrees of freedom) was obtained at $\zeta=47^\circ$, $\alpha=18^\circ$ 
(or vice versa) and $\mr=1.2$ ($g_r=0.80$). 
In Figure~1 we also show the lines of constant model pulsed fraction.
The lines for the pulsed fractions
compatible with the detected value, $f_p=30\pm 4\%$,
are close to the confidence contours
unless $\zeta$ and/or $\alpha$ are close to
$90^\circ$; at these large angles
the model light curves have a complicated shape (e.~g., two maxima 
per rotational period) inconsistent with what is observed.
The increase of the mass-to-radius ratio enhances
the gravitational bending
and leads to a greater contribution from the secondary PC (that on the
back NS hemisphere), which suppresses the model pulsations.
As a result, the allowed regions in Figure 1 completely vanish
at $\mr > 1.6$ ($g_r < 0.73$)\footnote{ 
In fact, at $\mr\gapr 1.93$ ($g_r \lapr 0.66$) the model pulsations
may grow because of appearance of strong narrow peaks from PC at
$\theta\simeq 180^\circ$. However, there are no such peaks in the observed
light curve.}.  With decreasing the $\mr$ ratio, the confidence regions
shift towards the bottom-left corner of the $\zeta$-$\alpha$ plane
reaching a limiting 
position at $\mr\simeq 0.3$ ($g_r\simeq 0.95$) when
the effect of the gravitational bending becomes negligible,
and the observer detects radiation only from the primary
PC (on the front hemisphere).

Figure~1 provides obvious constraints
on the pulsar mass-to-radius ratio. For instance, 
if there were no observational information about 
the $\zeta$ and $\alpha$ values, the only constraint
would be $M < 1.6\, \Ms\, (R/10~{\rm km})$, 
or $R>8.8\, (M/1.4 \Ms)$~km, at a 99\% confidence level.
If, however, we adopt $\zeta=40^\circ$ and $\alpha=35^\circ$,
as given by Manchester \& Johnston (1995), then
$1.4 < \mr <1.6$. Figure~2 shows the corresponding domain in the
NS mass-radius diagram,
restricted by the $M(R)$ dependences for soft ($\pi$) 
and hard (TI) equations of state of the superdense matter
(Shapiro \& Teukolsky 1983).
It follows from this picture, for example, that 
the radius of a NS with the canonical mass $M=1.4 \Ms$ is 
within the range $8.8<R<10.0$~km. 

Another set of angles, $\zeta=24^\circ$ and $\alpha=20^\circ$, was 
suggested by Gil \& Krawczyk (1997). This
set gets within the 99\% confidence region 
only for very low mass-to-radius ratios, 
$\mr < 0.3$, when the gravitational effects become negligible.
This corresponds to very low masses, $M < 0.5 \Ms$ at any $R$ 
allowed by the equations of state (Fig.~2). Note that for these angles and
$M_*/R_{10} < 1.8$ the secondary PC remains invisible during the whole
pulsar period.

\section{Conclusions}
We have demonstrated that 
the analysis of the soft X-ray radiation emitted by 
PCs of radio pulsars  
in terms of NS atmosphere models provides a new tool to constrain
the NS mass and radius, and consequently
the equation state of the superdense matter in the NS interiors. 
The constraints become more stringent if this analysis is
combined with complementary data on the
pulsar magnetic inclination and viewing angle, as we have shown using
the millisecond pulsar J0437--4715 as an example.
In principle, these angles can be inferred from the phase dependence
of the radio polarization position angle. However, in the case
of J0437--4715 this dependence is too complicated to be described
by a simple rotating vector model of Radhakrishnan \& Cooke (1969),
so that the inferred angles are very uncertain.
Once an adequate model for the radio emission is found,
the statistical analysis of the polarization data
would result in a domain of
allowed angles in the $\zeta$-$\alpha$ plane,
and the constraints should be based upon overlapping 
of the confidence regions obtained from  the X-ray light curve 
and from the radio polarization data. This may constrain
not only the $M/R$ ratio, but also the pulsar geometry.
 
The allowed $M(R)$ domain can be further restricted
if additional information on the NS mass is available.
For instance, since \psr is in a binary
system with a white dwarf companion,
it is possible to estimate independently an upper limit on the NS mass.
Sandhu et al.~(1997) found upper limits on the white dwarf mass,
$M_{\rm wd}\le 0.32\Ms$, and on the orbital inclination,
$i\le 43^\circ$, that yields a restriction $M < 2.5\Ms$ (see their Fig.~2). 
If further observations of the pulsar reduce the 
limit on $i$, or a lower upper limit on $M_{\rm wd}$ is obtained
from white dwarf cooling models, a more stringent constraint
on $M$ would follow, thus narrowing the allowed domains of the other
pulsar parameters ($R$, $\zeta$ and $\alpha$).

The above-described analysis for \psr~
 is simplified by the pulsar's low magnetic field
which does not affect the properties of the X-ray
radiation.  The spectra and, particularly, angular distribution of
radiation emerging from strongly magnetized NS atmospheres
depends significantly on the magnetic field (Pavlov et al.~1994).  Thus,
a similar, albeit more complicated, analysis of X-ray radiation from 
pulsars with strong magnetic fields, $B\sim 10^{11}-10^{13}$~G 
(e.~g., PSR B1929+10 and 0950+08),
would enable one to constrain also the magnetic field strength at their
magnetic poles.
\acknowledgments 
We thank Werner Becker for providing us with the
$ROSAT$ PSPC light curve.
We are grateful to Joachim Tr\"umper for stimulating discussions.
The work was partially supported through 
NASA grant NAG5-2807, INTAS grant 94-3834 and
DFG-RBRF grant 96-02-00177G. VEZ acknowledges the Max-Planck fellowship.  

\newpage
\figcaption{
68\%, 90\%, and 99\% confidence regions (dark, medium, and light
grey, respectively) in the $\zeta$-$\alpha$ plane for 
different values of the $\mr$ ratio. The lines with numbers 
show the values of constant pulsed fractions. The crosses mark 
the points with $\zeta=40^\circ$ and  $\alpha=35^\circ$.}

\figcaption{
NS mass-radius diagram with straight lines of constant $\mr$ ratio
(numbers near the lines) and $M(R)$ curves for
a few equations of state of superdense matter:
soft ($\pi$), intermediate (FP) and hard (TI and MF).
The line $R=1.5 R_g$ ($R_g$ is the gravitational radius) shows
the most conservative lower limit on the NS radius
at a given mass.
The hatched region shows the mass-radius domain 
for \psr~ allowed by the light curve fitting at 
$\zeta=40^\circ$ and  $\alpha=35^\circ$.}

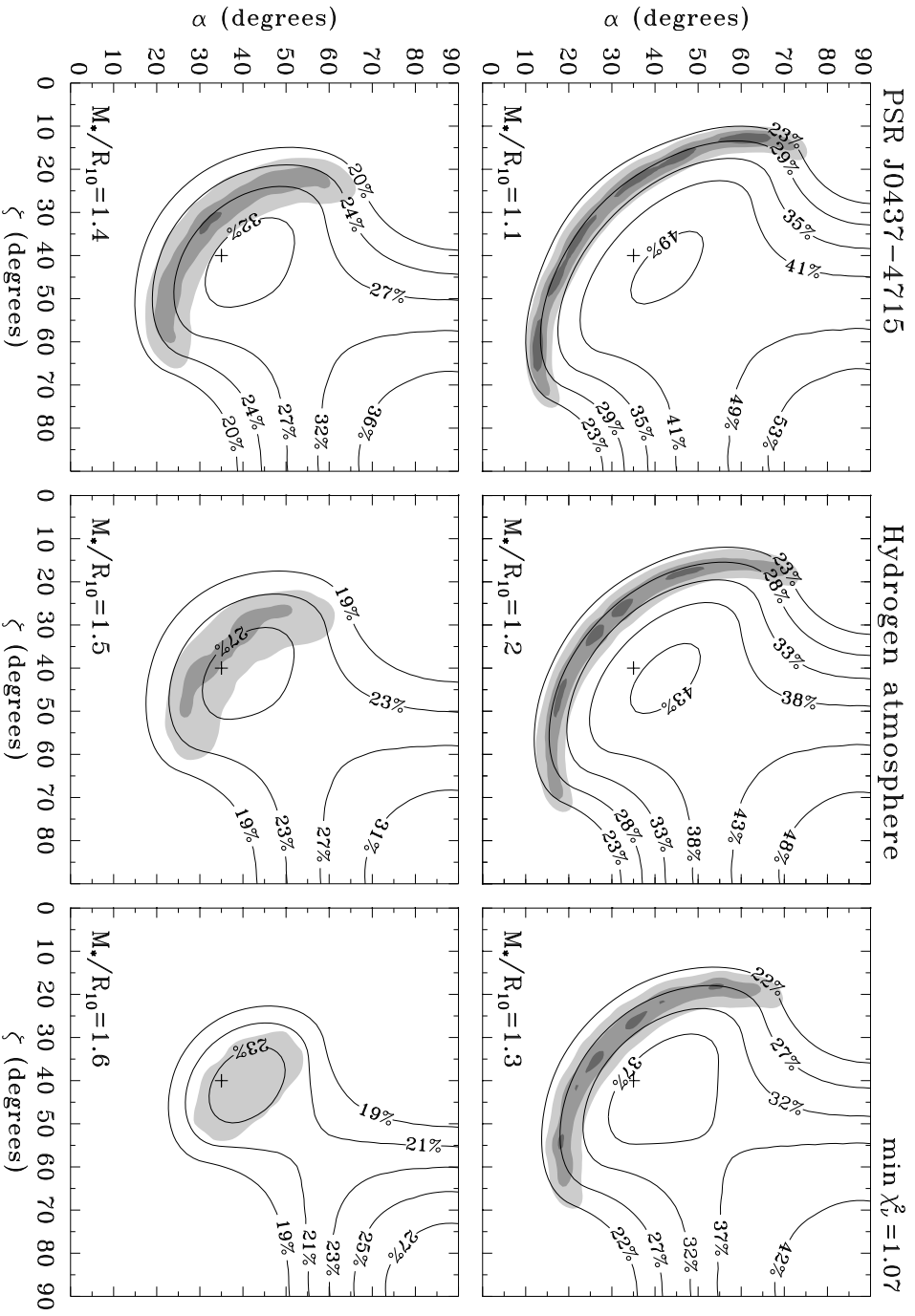
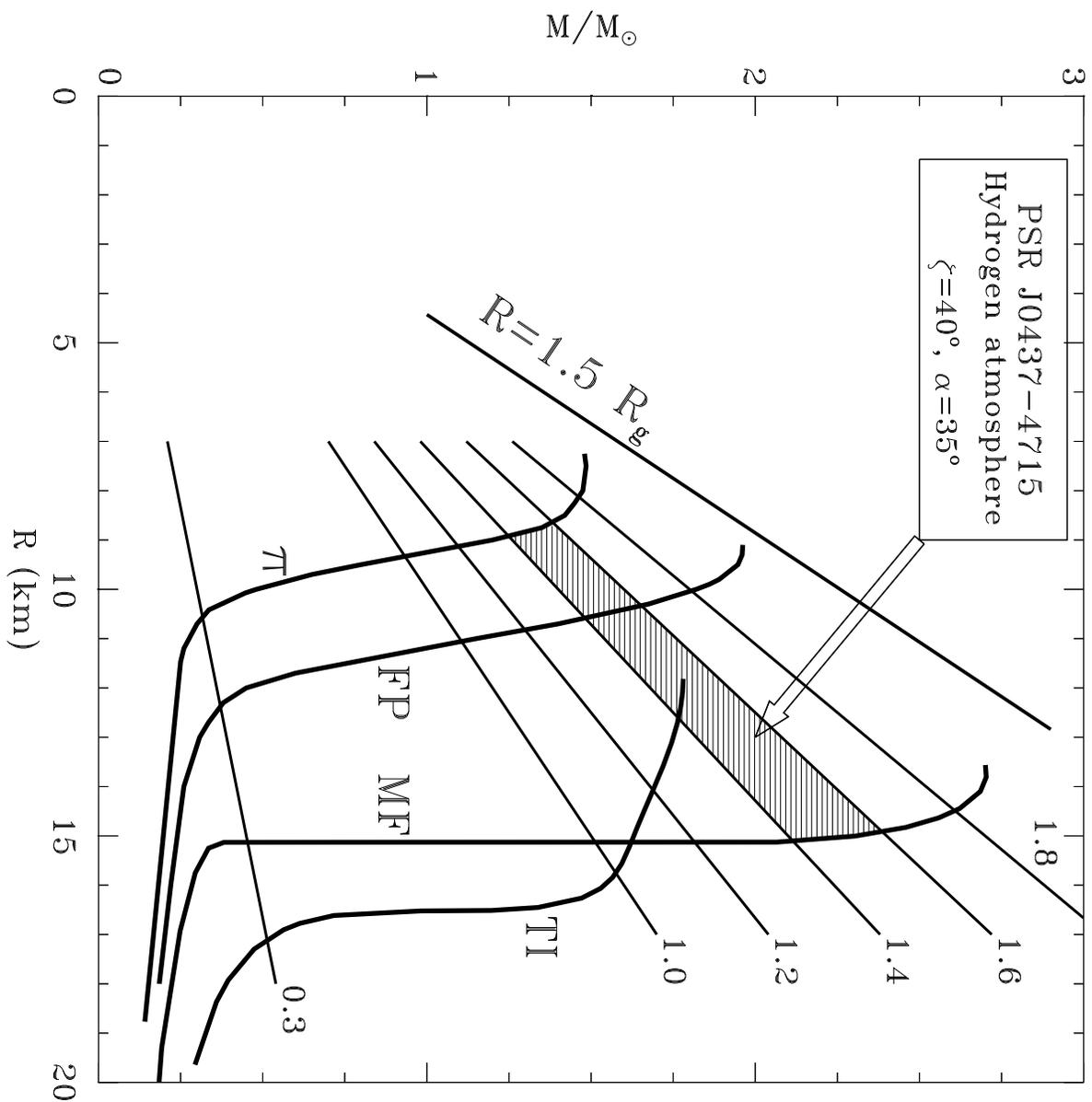
\end{document}